# Single point positioning using full and fractional pseudorange measurements from GPS and BDS

Sihao Zhao *, Xiaowei Cui and Mingquan Lu

Department of Electronic Engineering, Tsinghua University, No.1 Qinghuayuan, Haidian District, Beijing 100084, China

**Abstract:** In conventional global navigation satellite system (GNSS) receivers, usually full pseudorange measurements are required to complete a single point position fix. However, to obtain full pseudorange measurements takes longer time than for fractional pseudorange measurements. Considering such a fact, in order to shorten the time to first fix and improve the position accuracy during cold or warm start of a dual-constellation GNSS receiver, we propose a positioning algorithm using full and fractional pseudorange measurements from the two navigational constellations. This method uses four full pseudorange measurements from one constellation along with fractional ones from either or both constellations to obtain a potentially rapid position result with an identical accuracy to that of the conventional positioning method using full measurements. Tests with simulated and real Global Positioning System (GPS) and BeiDou Navigation Satellite System (BDS) data demonstrate that the proposed method can generate correct single point position solutions and the position error is identical with the result from the conventional approach using the full pseudorange measurements.

## 1. Introduction

For a conventional global navigation satellite system (GNSS) receiver, four full pseudorange measurements and valid ephemeris data are usually required to achieve single point positioning using one navigation constellation (Misra and Enge, 2006). A full pseudorange measurement can only be acquired after having obtained the second of week (SOW) information which is broadcast periodically by the navigation satellite and this takes at least 6 seconds for the Global Positioning System (GPS) as an example. In order to shorten the time to first fix under receiver cold or warm start conditions and increase the receiver's positioning robustness when fewer full pseudorange measurements are observed, sub-millisecond or sub-twenty-millisecond fractional pseudorange measurements obtained before the completion of frame synchronization are considered to be adopted in positioning. Approaches presented in (van Diggelen, 2009) and (Akopian and Syrjarinne, 2009) can be used to complete single point positioning after recovery of the full pseudorange measurements as long as valid ephemeris data and an acceptable initial position estimate are given. An exhaustive grid searching method which requires proper initial estimation boundaries is proposed and validated through experiments using simulated GPS data (Sirola, 2006). Another coarse-time GPS positioning algorithm is discussed using the constraints of satellite distances to reduce search complexity (Jing et al., 2017b). Techniques that introduce Doppler measurements combined with pseudorange ones from one navigation constellation are also proposed and tested to instantaneously solve the receiver position results (Chen et al., 2014, Fernández-Hernández and Borre, 2016). As a new comer to GNSS, BeiDou Navigation Satellite System (BDS) is being adopted by more positioning applications in recent years. In (Jing et al., 2015), Doppler measurements are used to help recover the full pseudorange measurements of a BDS receiver and obtain a single point position solution. Altitude constraints are also adopted to assist BDS receivers to shorten the time to first position fix (Jing et al., 2017a). BDS has some unique features other than GPS based on which (Zhao et al., 2016) proposes a single point positioning method using at least four full pseudorange measurements from BDS geostationary satellites (GEOs) and at least one fractional measurement from non-GEOs which is independent of initial position estimate. The studies mentioned above ideally can shorten the time to first fix from more than 6 s to less than 1 s. From the aspect of accuracy, these methods can output single point positioning results during the time when conventional approaches using only full pseudorange measurements are unable to give even a position fix and thus improve the overall accuracy of the receiver throughout its working time. However, these studies are all focused on the single constellation case, and whether fractional pseudorange measurements can be used for the cases with two constellations such as GPS and BDS to improve positioning accuracy and speed has not been investigated.

Nowadays, multi-constellation processing has become a standard configuration for GNSS receivers, e.g. BDS+GPS dual-constellation receivers. Combining GPS and BDS measurements, either pseudoranges or carrier phases or both, to achieve higher positioning accuracy and robustness attracts attention from researchers and has been reported in several previous studies, such as (Kong et al., 2016, Zhao et al., 2014, Odolinski and Teunissen, 2016, Teunissen et al., 2014), to name a few. However, the pseudorange measurements used in these studies are all the so-called full pseudorange measurements, and how the fractional parts of the measurements can be utilized in positioning is not discussed. Considering that full measurements from one of the constellations and fractional measurements from either BDS or GPS or both are adopted in positioning, the receiver's performance such as time to first fix, positioning accuracy and robustness can be potentially improved. For example, if a receiver has already obtained four or more full pseudorange measurements from one constellation, the adoption of fractional measurements which can possibly be used as full measurements, from either constellation can usually help improve positioning accuracy and robustness. Based on this motivation, a single point positioning method for the dual-constellation case using full measurements from one constellation and fractional measurements from the other or both constellations is proposed and verified.

In the following text, first, the measurement model of the full and fractional measurements from two navigation constellations is presented. Then, based on the characteristic of the inter-system time offset between the two navigation constellations, a mixed single point positioning algorithm for the dual-constellation case is proposed, which adopts fractional and full measurements from two constellations such as GPS and BDS.

The procedure of the method and its successful condition are discussed as well. After that, the proposed method is validated using simulated and field collected BDS and GPS measurement data. Lastly, a conclusion is drawn and possible future work is briefly discussed.

## 2. Model of Full and Fractional Pseudoranges from Dual Constellations

The pseudorange measurements from two navigation constellation should be modelled first. The two constellations are denoted as $A$ and $B$, respectively. Without loss of generality, the full pseudorange measurements are assumed to be from constellation $A$, and the fractional measurements from either or both $A$ and $B$ constellations. The full measurements of constellation $A$ are given by (1).

$$z_i = \left|\boldsymbol{X}_i - \boldsymbol{x}_r\right| + b + \varepsilon_i \quad (1)$$

where $z_i$ is the full pseudorange measurement (unit: m), $i$=1,2,…,$n$ is the index of the full-measurement satellites, $\boldsymbol{X}_i$ is the position vector of the full-measurement satellite, $\boldsymbol{x}_r$ represents the receiver position, $b$ is the receiver clock bias (unit: m) with respect to $A$, $\varepsilon$ represents other measurement errors and noises.

The fractional measurements of $A$ are modelled as (2).

$$z_j = \left|\boldsymbol{X}_j - \boldsymbol{x}_r\right| + b - N_j c_T + \varepsilon_j \quad (2)$$

where, $z_j$ is the fractional measurement, $\boldsymbol{X}_j$ is the position of the fractional-measurement satellite, $N_j$ is the integer part of the fractional measurement, i.e. the remainder after dividing the corresponding full measurement by $c_T$, $c_T$ (unit: m) is the traveling distance of light within $T$, $T$ is a constant time length (unit: s), e.g. $T$=1 ms for 1 ms period pseudorandom code synchronization or 20 ms for 20 ms data bit synchronization in the case of GPS L1 civilian signal, $j$=$n$+1,$n$+2,…,$n$+$n_A$ is the index of the fractional-measurement satellites from $A$.

The fractional measurements of $B$ are a bit different from that of $A$ due to the system time offset between the two constellations. The system time of $A$ is selected as the reference and thus the fractional measurement model of $B$ is given by

$$z_j = \left|\boldsymbol{X}_j - \boldsymbol{x}_r\right| + b + b_{AB} - N_j c_T + \varepsilon_j \quad (3)$$

where $b_{AB}$ is the time offset between $A$ and B, and $j$=$n$+$n_A$+1,…,$n$+$n_A$+$n_B$ is the index of the fractional measurement satellites from $B$.

## 3. Positioning with Mixed Pseudorange Measurement from Dual Constellations

The unknowns to be solved in (1), (2) and (3) include the receiver position, receiver clock bias, system clock offset and all integer ambiguities of the fractional measurements, as given by

$$\boldsymbol{X} = \left[\boldsymbol{x}^T, \boldsymbol{N}^T\right]^T \tag{4}$$

where $\boldsymbol{x} = \left[x, y, z, b, b_{AB}\right]^T$ and $\boldsymbol{N} = \left[N_1, N_2 \cdots, N_{n_A+n_B}\right]^T$.

Compared with the single constellation case (Zhao et al., 2016), besides the receiver position, clock bias and pseudorange integer ambiguities, the unknowns have an extra time offset $b_{AB}$. The amount of the unknowns exceeds the amount of equations which means a unique solution is impossible to be obtained.

However, the integer feature of the ambiguity $N$ can provide some useful constraints: if $b_{AB}$ is sufficiently small that does not lead a carry to $N$, then $b_{AB}$ can be temporarily omitted during computation. After the integer $N$ is correctly calculated, the full pseudorange measurements of $B$ can be recovered using (5).

$$z_{j,full} = z_j + N_j c_T \tag{5}$$

At this stage, $b_{AB}$ is then treated as an unknown as given by (6). Conventional dual-constellation single point positioning approach (Borre et al., 2007) can be used to estimate the receiver position and time offset $b_{AB}$.

$$\boldsymbol{X}_{new} = \left[x, y, z, b, b_{AB}\right]^T \tag{6}$$

Furthermore, the conditions to correctly recover the ambiguities should be discussed. Similar to the single constellation case, the correct integer ambiguity can be rounded correctly as long as the estimated range error of the fractional-measurement satellite is smaller than half code or half bit length. One thing to note is that this estimated range error is not only caused by the full measurements but also brought by the time offset between systems. Therefore, the constraint to correctly recover the integer ambiguities is written as follows,

$$\left|\sum_{i=1}^{n}\left(\delta\rho_i \boldsymbol{e}_i^T \boldsymbol{e}_j\right) + b_{AB}\right| < \alpha \tag{7}$$

where $\delta\rho_i$ is the range error of the $i$th full-measurement satellite, $\boldsymbol{e}_i$ represents the normalized line-of-sight (LOS) vector from the user to the full-measurement satellite $i$, $\boldsymbol{e}_j$ represents the normalized LOS vector from the user to the fractional-measurement satellite $j$, and $\alpha$ is the half cycle threshold, e.g. ~150 km for the half coarse acquisition (C/A) code-length case.

Moreover, in order to simplify the expression, the geometric dilution of precision (GDOP) of all the full-measurement satellites is introduced as

$$\left|\sum_{i=1}^{n}\left(\delta\rho_i \boldsymbol{e}_i^T \boldsymbol{e}_j\right)+b_{AB}\right| \leq \left|\sum_{i=1}^{n}\delta\rho_i \boldsymbol{e}_i^T\right|+\left|b_{AB}\right|=\left|\delta \boldsymbol{P}\right|+\left|b_{AB}\right| \leq \text{GDOP}\times \max_{i\in\{1,2,\cdots,n\}}\left(\left|\delta\rho_i\right|\right)+\left|b_{AB}\right| \quad (8)$$

where,

$$\text{GDOP}=\sqrt{trace\left(\left(\begin{bmatrix}-\boldsymbol{e}_1 & \cdots & -\boldsymbol{e}_n \\ 1 & \cdots & 1\end{bmatrix}\begin{bmatrix}-\boldsymbol{e}_1^T & 1 \\ \vdots & \vdots \\ -\boldsymbol{e}_n^T & 1\end{bmatrix}\right)^{-1}\right)}=\sqrt{\sum_{i=1}^{4}\frac{1}{\lambda_i}}$$

and $\lambda$ is the eigenvalue of the matrix in the brackets.

Consequently, the condition to successfully resolve the pseudorange integer ambiguity is given by

$$\text{GDOP}=\sqrt{\sum_{i=1}^{4}\frac{1}{\lambda_i}} < \beta=\frac{\alpha-\left|b_{AB}\right|}{\max_{i\in\{1,2,\cdots,n\}}\left(\left|\delta\rho_i\right|\right)} \quad (9)$$

In the BDS and GPS dual-constellation case, the time offset between GPS and Universal Time Coordinated (UTC) is within 1 us (Directorate, 2013), and the offset between BDS and UTC is within 100 ns (CSNO, 2013b). Therefore, the time offset between GPS and BDS is at most 1.1 us or 330 m. The $3\sigma$ user equivalent range error (UERE) of both GPS and BDS is assumed to be less than 50 m which is a conservative estimate according to (CSNO, 2013a, Kaplan and Hegarty, 2006). As a result, $\beta$ can be set to (150000-330)/50=2993.4 for BDS + GPS applications. Figure 1 depicts the process of the dual-constellation mixed full- and fractional- measurement positioning algorithm.

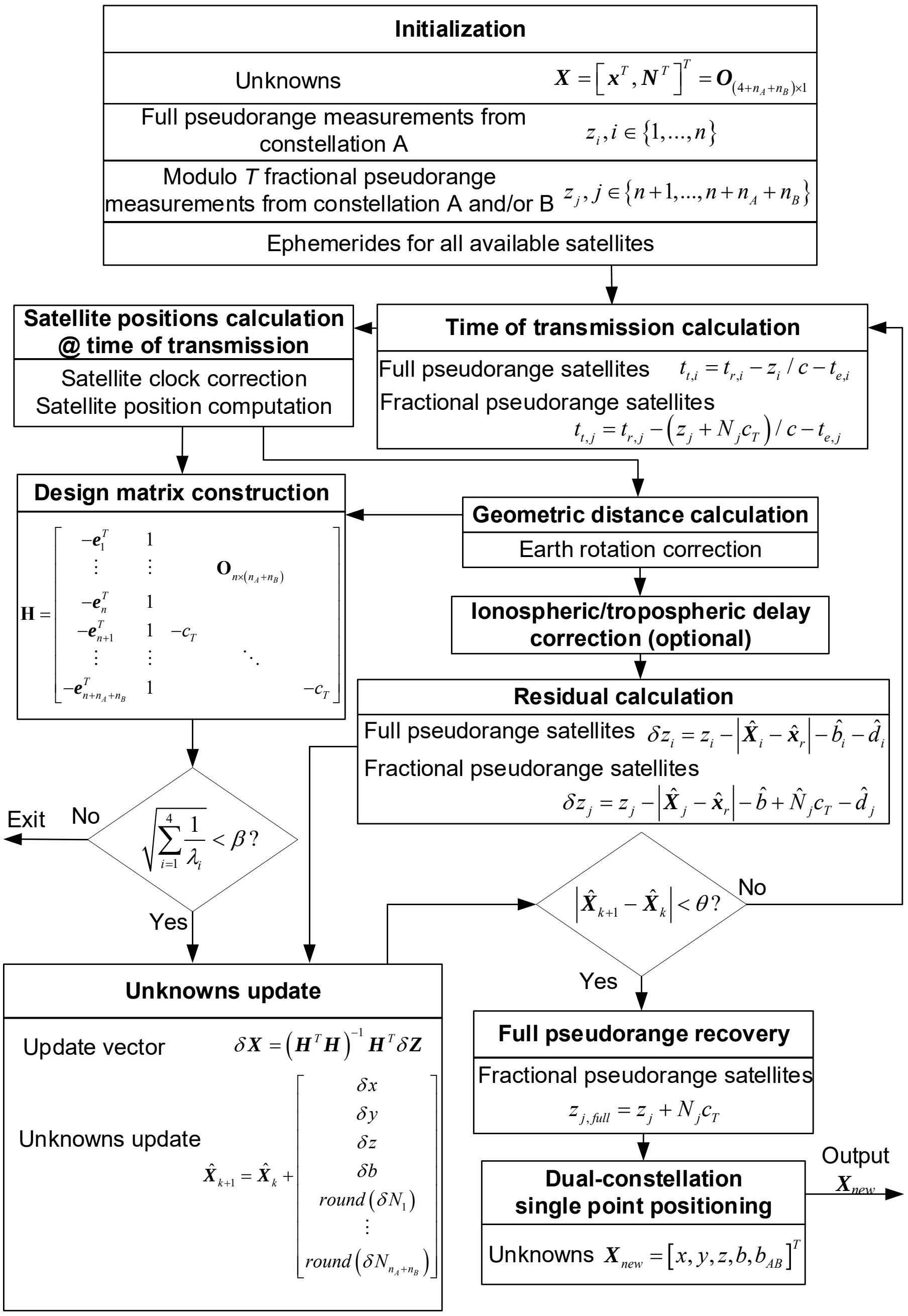

**Figure 1.** Flowchart of the full and pseudorange measurement mixed dual-constellation positioning algorithm

From Figure 1, it can be seen that in the initialization step, we omit the time offset between the two constellations temporarily and list only the three axis coordinates and the time bias of the user receiver as unknowns. The ephemeris data from the two constellations should also be ready for satellite position computation. Then the algorithm enters a large iteration step in which satellite positions are computed for a) design matrix construction and b) earth rotation and path error correction. The eigenvalue metric given by (9) is applied to test if the algorithm can enter the next step. Measurement residuals are calculated as shown in the residual calculation module. Then the incremental vector for the unknowns are generated using the design matrix and the residuals, to update the user position and pseudorange integer ambiguities. The iteration runs until the norm of the incremental vector is sufficiently small. Afterwards, the corresponding full measurements of the fractional pseudoranges are recovered. Finally, introducing the constellation time offset back to the unknowns, the user position result is obtained using a conventional dual-constellation single point positioning approach.

## 4. Experiments and results

In the following experiments, a simulation test is firstly conducted to verify the proposed dual-constellation single point positioning algorithm in an extreme case. To further examine the performance under real environments, field-collected real data at three different places are used to test the pseudorange recovery and the positioning accuracy of the algorithm.

In order to validate the proposed algorithm, the prerequisite of at least 4 full pseudorange measurements from either constellation should be satisfied. Besides, the eigenvalue test in the algorithm requires the GDOP of the full-pseudorange satellites smaller than a threshold to proceed position computation. We notice that in the current GNSS satellites in space, BDS has GEO satellites that broadcast faster navigation data which makes it easy to obtain their full pseudoranges. The navigation data rate of BDS GEOs is 500 bps which is 10 times as high as that of the BDS non-GEOs and GPS(Directorate, 2013, CSNO, 2013b). The BDS GEO sub frame has 300 bits and contains one SOW in each, which means it is possible to obtain a full pseudorange measurement within 0.6 s theoretically. Moreover, the GEOs also have chance to form a bad geometry and an enormous GDOP value can be observed at such times which is suitable to test the feasibility of the proposed eigenvalue criterion. Therefore, BDS is selected as one of the two constellations in both following simulation and real data experiments. Full pseudorange measurements from BDS GEOs and other measurements from non-GEOs are used to test the proposed algorithm. GPS is selected as another constellation because it is widely used and the orbital data are simple to obtain so that the simulation can be implemented easily.

### *4.1. Simulation Test*

*Methodology and Data*

The real ephemeris data of BDS and GPS on May 19, 2015 are used to simulate the two constellations. A snapshot of the constellations is depicted in Figure 2.

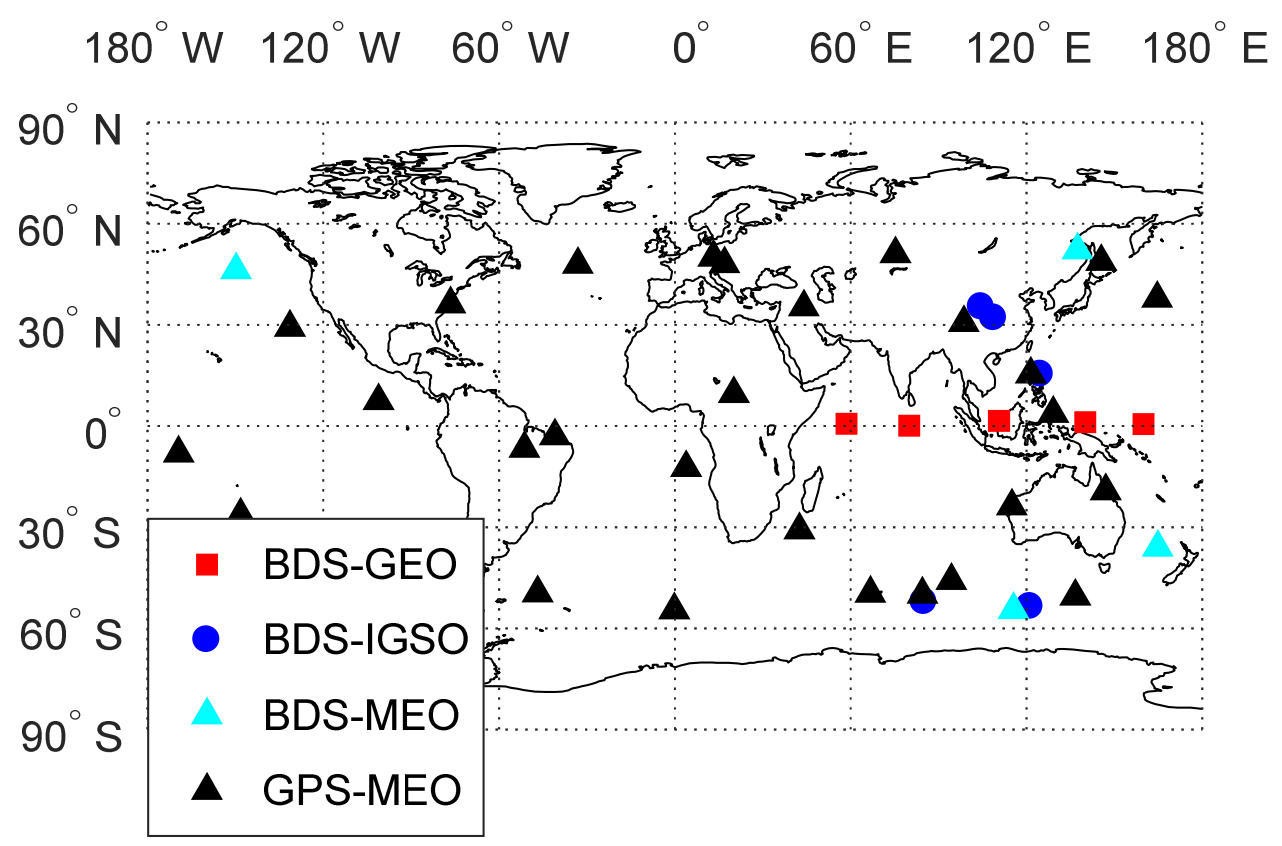

**Figure 2.** Snapshot of the BDS and GPS constellations in the simulation

The earth surface is sampled with a step of one degree along latitude and longitude. The GNSS receiver is assumed to be placed on these 1-degree-distance ground points. With a time step of 30 minutes, eight-day epoch-by-epoch satellite positions are simulated. At each epoch, the visibility of the BDS GEOs and their GDOP observed from all the ground points are calculated based on the simulated satellite positions. At most epochs, if a receiver at a selected ground point can see four or more GEOs, the GEO GDOP is smaller than the threshold, i.e. 2993.4. However, there are cases that at some epochs, some ground points cannot obtain a small GDOP even if four or more GEOs are visible. Among them, the worst epoch, when the amount of ground points or the ground area that have a GDOP smaller than the threshold is minimal, is selected for further investigation, as depicted in Figure 3. In this case, the ground area where GDOP is smaller than the threshold, i.e., area inside the red dash-dot contour in Figure 3, occupies only 76.42% of the area where 4 or more GEOs are visible (blue full-line contour in Figure 3). This is an extreme case that can help examine if the eigenvalue test of the algorithm works, as well as test the performance of pseudorange recovery and single point positioning. In other words, the proposed algorithm is expected to yield correct position results inside the red dash-dot area but report eigenvalue test failure outside the red dash-dot but inside the blue full-line contour.

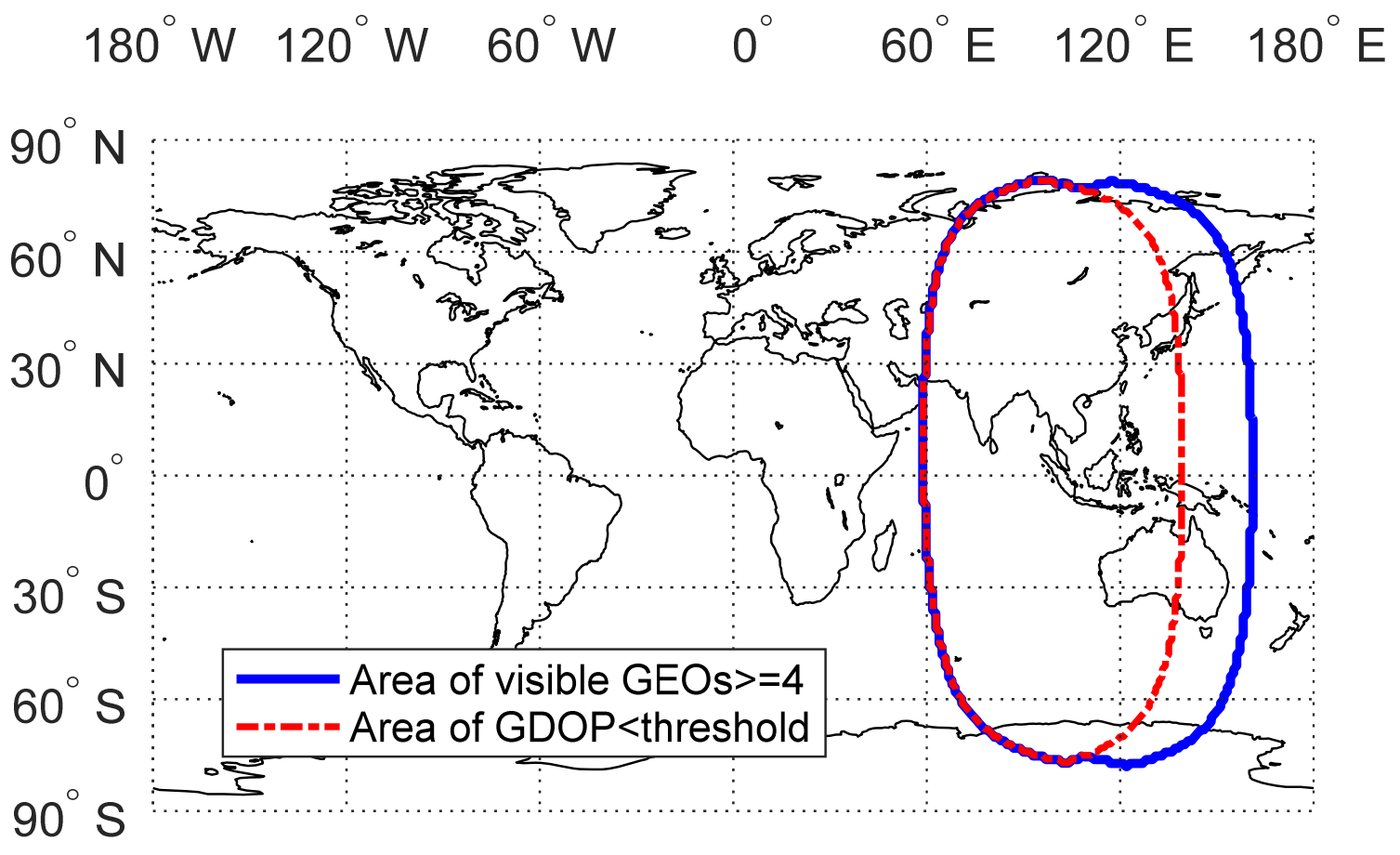

**Figure 3.** Worst epoch when the ground area with GEO GDOP<2993.4 is minimal

At a specific time epoch, all the ground points are traversed to calculate the real distances between all the simulated visible satellites and the ground receiver. Then these theoretical ranges are added with a clock bias and measurement errors to form the simulated full pseudorange measurements as given by (10).

$$PR_i = D_i + b + \varepsilon_i \tag{10}$$

where $PR_i$ is the full pseudorange measurement to the satellite $i$, $D_i$ is the receiver-to-satellite distance, $b$ is the clock bias between the receiver and the constellation, and $\varepsilon$ is the measurement error.

The simulated full pseudorange measurements of the non-GEOs are then truncated to remain their sub-millisecond parts to form the fractional measurements, and the simulated GEO measurements remains as is. These fractional and full measurements compose the input to the proposed algorithm. The original full measurements of all satellites can be used to compute the receiver position as a comparison. The offset between BDS and GPS is set to the upper limit of 1.1 μs as analyzed in the previous section. The actual clock bias between the receiver and the BDS time is set to 10 ms, and thus the clock bias between the receiver and GPS time is 0.0100011 s. The pseudorange measurement errors for both BDS and GPS are set to 1.3 m (1-sigma). The initial position and clock bias estimates of the receiver are all set to zero.

*Simulation Test Result*

After the simulated full and fractional pseudorange measurements at all ground points at the worst epoch are processed by the proposed algorithm, the area where the algorithm can generate correct position results are shown as the red full-line contour in Figure 4. The area where the algorithm reports failure in the eigenvalue test is depicted as the blue dash-dot contour in the same figure. After inspection into all the ground points in these two areas, it can be found that the area of successful position results are identical with the area with a GDOP smaller than the threshold, i.e. red dash-dot contour in Figure 3. Besides, the area where eigenvalue test fails is identical with the area where the amount of visible GEOs is 4 or more but the GDOP exceeds the threshold. These indicate that the algorithm can work successfully when at least four full pseudorange measurements are obtained and the eigenvalue test can correctly identify if the algorithm can continue to compute the correct position result or not.

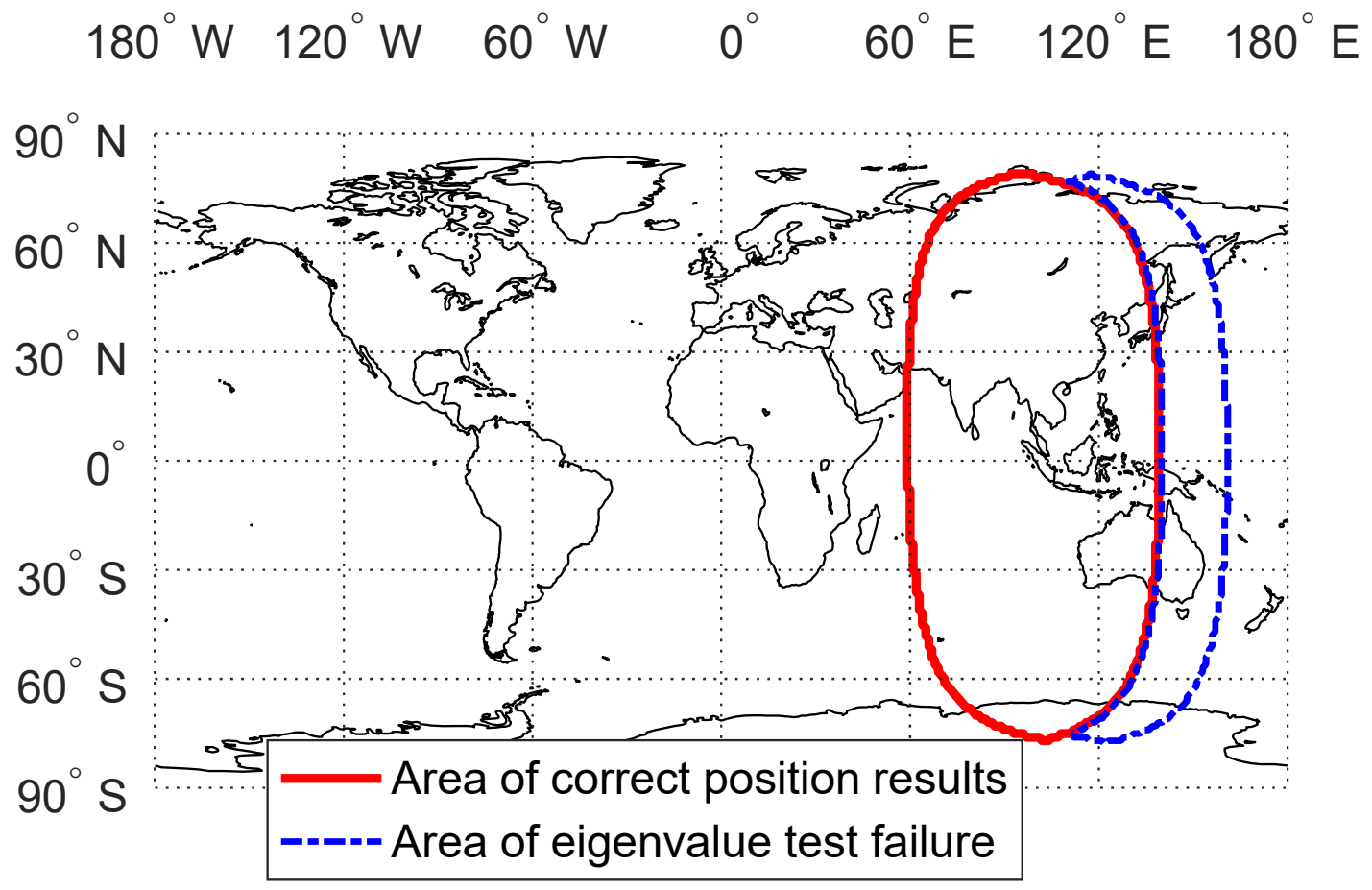

**Figure 4.** Area in which correct position results are obtained (red contour), and area with eigenvalue test failure (blue dash-dot contour)

As described in Section 3, full pseudoranges can be recovered from the fractional ones by the proposed algorithm. Once the full pseudoranges are correctly calculated, the positioning computation can be done. Therefore, it is necessary to check if this intermediate variable is corrected obtained by the proposed approach. The recovered full pseudorange results are shown in Table 1. Eight non-GEOs instead of all satellites are selected to save space. The second column shows their initial simulated full pseudorange measurements, the third column shows their corresponding fractional values after modulo by 1 ms light travel distance, and the last column contains the full measurements recovered by the proposed algorithm. It can be seen that they are all correctly recovered. Moreover, this means the fractional measurements can be used as full ones to participate into positioning, which can usually improve positioning speed, accuracy and robustness.

**Table 1.** Recovered full pseudorange result from the fractional pseudoranges by the proposed algorithm (simulation)

| Satellite No. | Real full pseudorange measurement /m | Chopped fractional measurement /m | Recovered full pseudorange /m |
|---|---|---|---|
| BDS 6 | 42578331.90 | 7802.87 | 42578331.90 |
| BDS 7 | 43977853.55 | -91637.78 | 43977853.55 |
| BDS 13 | 24920128.79 | 37354.78 | 24920128.79 |
| BDS 14 | 26122983.62 | 41039.78 | 26122983.62 |
| GPS 3 | 28345069.65 | -135213.86 | 28345069.65 |
| GPS 11 | 26047901.81 | -34042.04 | 26047901.81 |
| GPS 18 | 23974396.63 | -9000.01 | 23974396.63 |
| GPS 26 | 24044662.90 | 61266.27 | 24044662.90 |

The position results at all the ground points inside the red contour in Figure 4 given by the proposed algorithm are illustrated in the left three curves of Figure 5. The x-axis represents the ground point index, the three y axes represent the three respective components of the positioning result in the earth-centered-earth-fixed (ECEF) coordinate. As a comparison, the three curves on the right side are generated by the conventional dual-constellation single point positioning method using all available full pseudorange measurements. Both methods have identical position errors, which indicates that the proposed algorithm can successfully compute the user position with an identical accuracy compared with the conventional full measurement method. This also indicates the correct computation of the recovered full pseudoranges which is shown in Table 1.

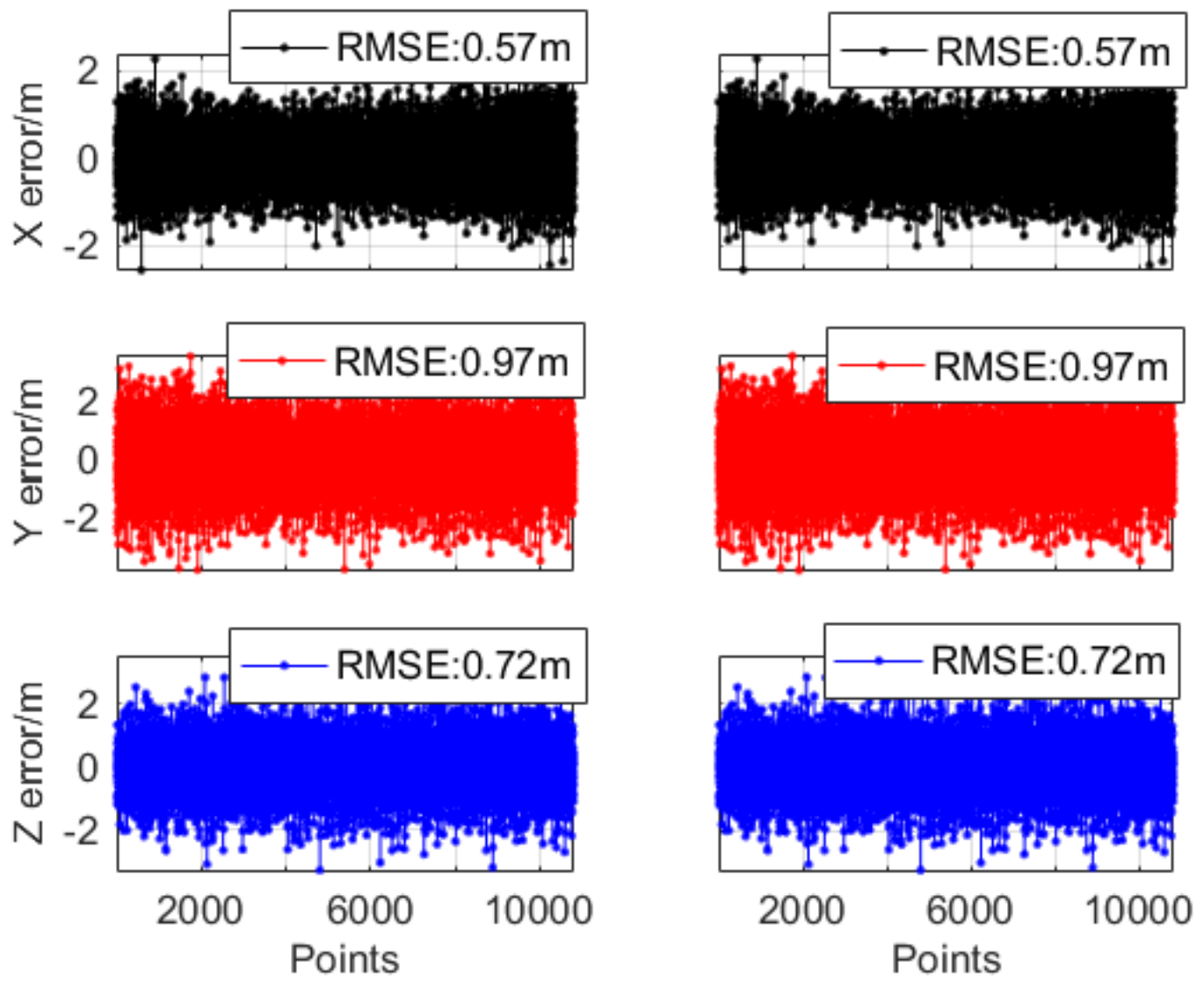

**Figure 5.** Position result comparison of the mixed method (left column) and the conventional dual-constellation method (right column) using simulated data

### *4.2. Real Data Experiment*

*Methodology and Data*

To verify the proposed positioning algorithm in real world, real BDS and GPS measurement datasets at three different places are collected, one from a self-developed BDS+GPS receiver, and two from the International GNSS Service (IGS) BKG data center (BKG, 2018). The first dataset was recorded by a self-developed receiver during 3:46 and 6:46 (UTC), Nov-21, 2015, and static GPS and BDS data as well as the broadcast ephemeris data were stored on the rooftop of Weiqing Building, Tsinghua University, Beijing, China. The second dataset contains the observation data during 0:00 Oct-1 and 0:00 Oct-4, 2018 (UTC) from the IGS Site AMNG in Putrajaya, Malaysia, and the broadcast ephemerides during the same period. The third dataset recorded the data from 0:00 Oct-1 to 0:00 Oct-4, 2018 (UTC) from IGS Site TOW2, Cape Ferguson, Australia, and the corresponding broadcast ephemeris data. The geographic locations where the three datasets are collected are depicted in Figure 6.

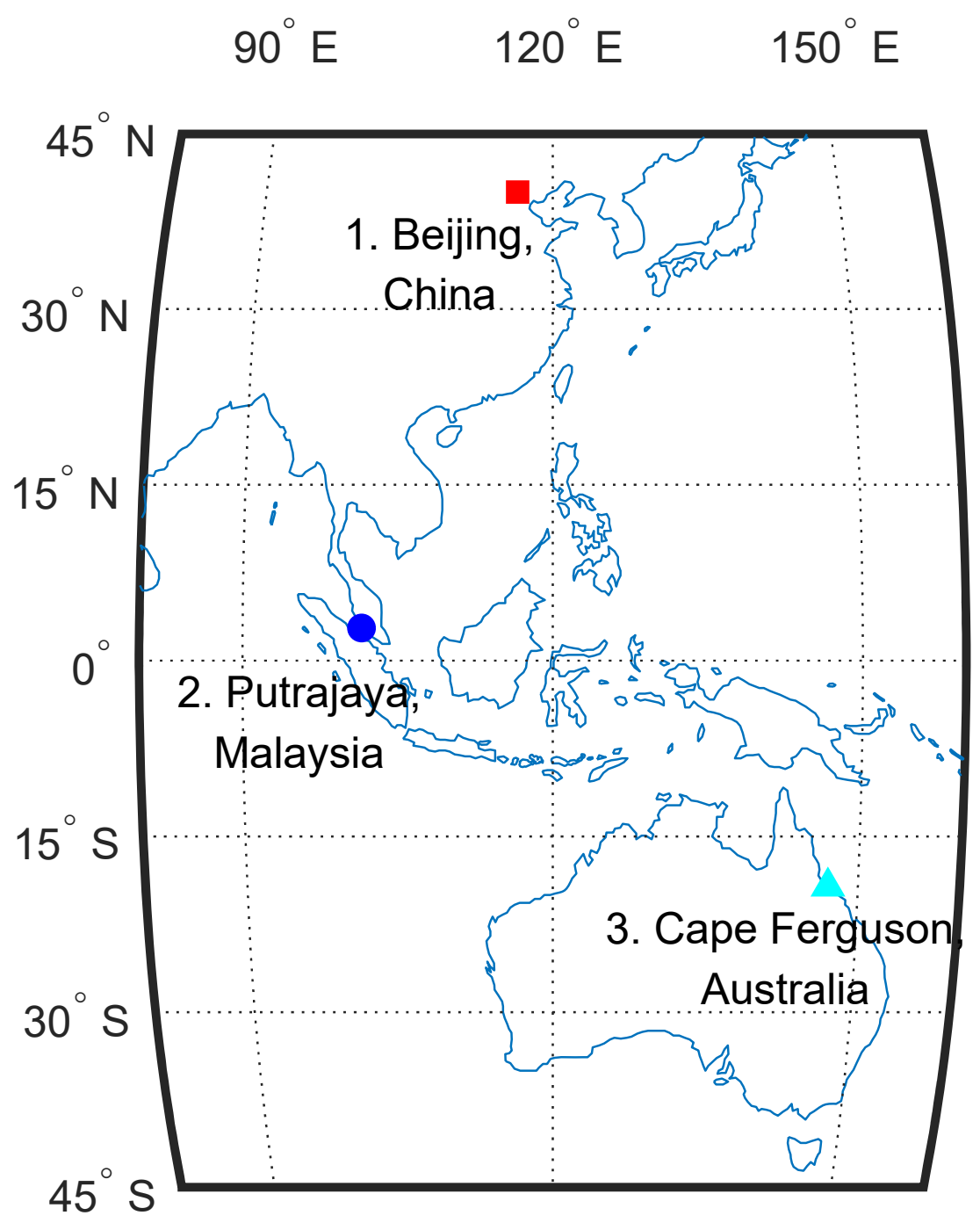

**Figure 6.** Locations where the three real-world datasets were generated

According to the principles of the proposed algorithm described above, at least four full pseudorange measurements are required. For BDS receivers, the GEO full measurements can be obtained earlier due to the fast broadcast data rate while only fractional measurements can be obtained for other satellites of BDS and GPS. In order to use the fractional pseudorange measurements to verify the algorithm, the field-collected full measurements of BDS non-GEOs and GPS satellites are chopped, i.e., modulo 1 ms fractional remainders are obtained, to simulate the code synchronization case. Examples of the real full pseudorange measurements and their chopped counterparts are shown in the second and the third column of Table 2, respectively.

*Real Data Test Result*

The position results generated by the proposed algorithm using the three datasets are demonstrated and discussed in the following texts.

Dataset #1

The last column of Table 2 is the recovered full pseudoranges by the proposed approach for dataset #1 collected in Beijing, which are identical to the original ones and shows the correctness of the method.

**Table 2.** Recovered full pseudorange result from the fractional pseudoranges by the proposed algorithm (real data #1)

| Satellite No. | Real full pseudorange measurement /m | Chopped fractional measurement /m | Recovered full pseudorange /m |
|---|---|---|---|
| BDS 6 | 36287456.97 | 12569.55 | 36287456.97 |
| BDS 8 | 36609266.69 | 34586.82 | 36609266.69 |
| BDS 9 | 37460125.38 | -13931.87 | 37460125.38 |
| GPS 2 | 20437145.56 | 51258.42 | 20437145.56 |
| GPS 5 | 20758476.18 | 72796.58 | 20758476.18 |
| GPS 6 | 21576684.93 | -8372.05 | 21576684.93 |
| GPS 9 | 22827710.52 | 43483.71 | 22827710.52 |
| GPS 12 | 23675017.99 | -8586.19 | 23675017.99 |
| GPS 13 | 24646443.41 | 63461.85 | 24646443.41 |
| GPS 25 | 23891094.88 | -92301.76 | 23891094.88 |
| GPS 29 | 24073806.02 | 90409.38 | 24073806.02 |

Moreover, the positioning results using the proposed method are demonstrated in Figure 7, and the results of the conventional dual-constellation positioning method are included as a comparison. It can be seen that the proposed method outputs the correct position results that have an identical accuracy with the results from the conventional approach using all full measurements. This also indicates a correct estimate and successful recovery of the full pseudoranges for the fractional measurements.

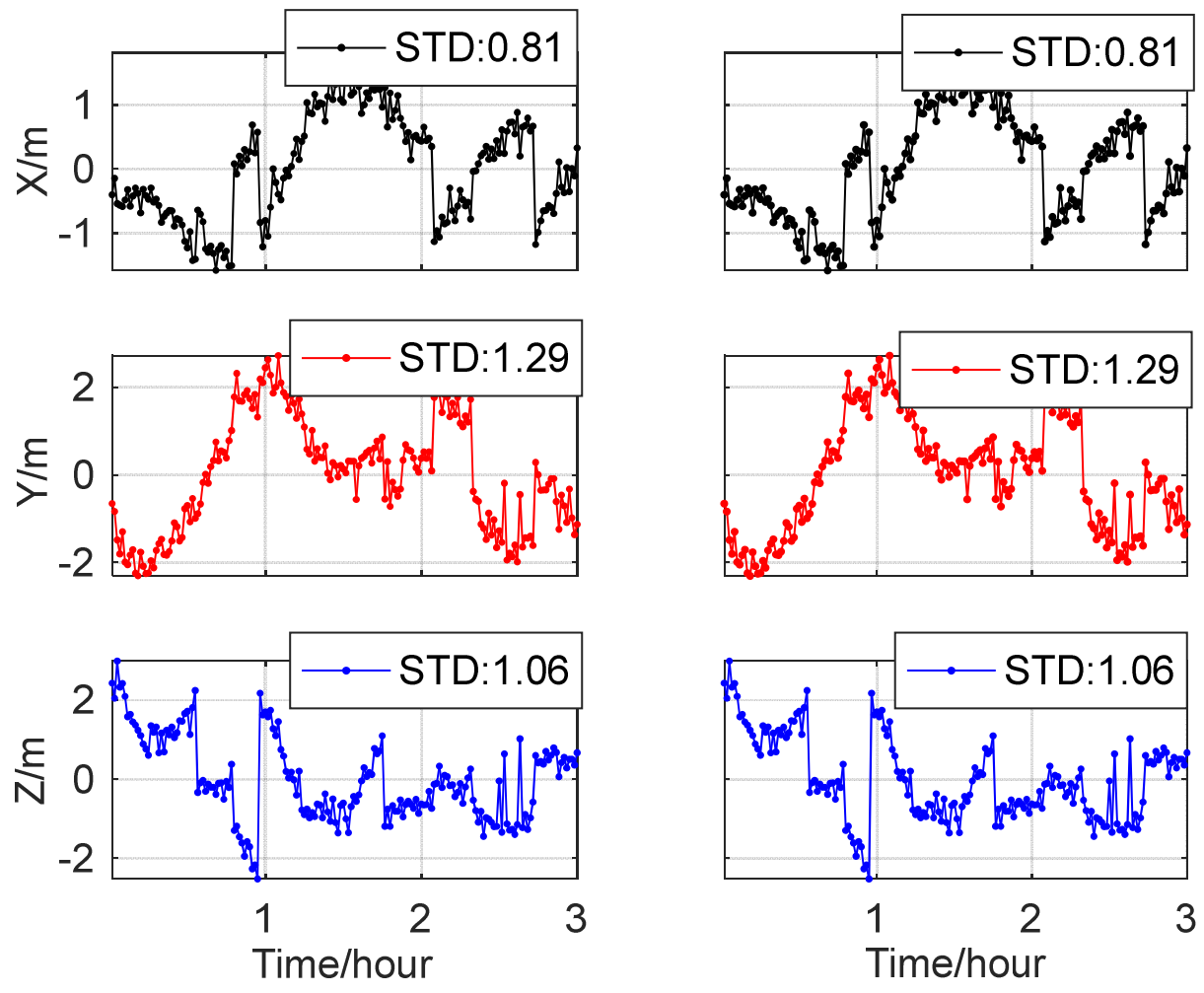

**Figure 7.** Position result comparison of the proposed mixed method (left column) and the conventional dual-constellation method (right column) using real data from dataset #1

Datasets #2 and #3

Another two datasets from other two locations containing 72 hours of measurements data are used to further demonstrate the performance of the proposed method in processing longer time length of data at different locations. The results in Figure 8 show that the position error of the proposed algorithm is identical with that of the conventional approach in different places of the world. To save space, the pseudorange measurement recovery result which is similar to that of dataset #1 is not listed here. From dataset #3, we observe that the proposed algorithm reports that 4.98% out of the total measurement epochs do not pass the eigenvalue test due to the poor geometry of the full-measurement satellites, i.e., the BDS GEOs, at the measurement time. This verifies the correctness and effectiveness of the eigenvalue metric test which increases the robustness of the proposed algorithm in extreme cases.

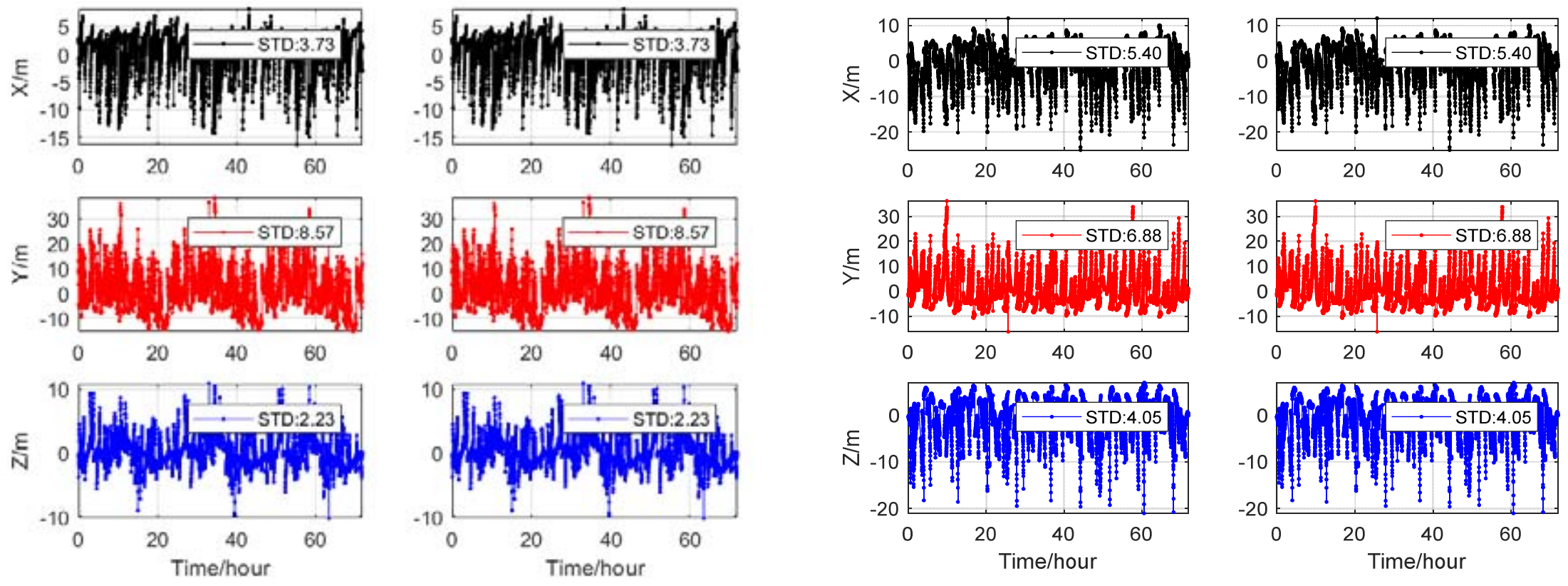

(a) Dataset #2 result (b) Dataset #3 result

**Figure 8.** Position result comparison of the proposed mixed method (left columns of (a) and (b)) and the conventional dual-constellation method (right columns of (a) and (b)) using real data from datasets #2 and #3

## 5. Conclusion and Future Work

A single point positioning approach using full and fractional pseudorange measurements for the double-constellation case is proposed to shorten the time to first fix and improve the position accuracy before full pseudorange measurements from all visible satellites are obtained. Four or more full measurements from one constellation are used to correctly recover the full measurements of the fractional counterparts from both constellations, as long as the time offset between the two constellations is smaller than a threshold as given in this work and the ephemeris data are valid. The conditions under which the proposed method can successfully generate the correct position results are analysed, and the judgment threshold is presented analytically with an example based on the officially announced performance of BDS and GPS. The procedure of the proposed algorithm is also presented in detail.

Simulated and real field-collected data are used to validate the proposed approach and the results show that the positioning solutions using the mixed full and fractional measurements are identical with that from the conventional method using all full pseudoranges. The proposed algorithm can potentially shorten the time to first fix after a receiver restart based on its principle of using fractional measurements obtained earlier than the full pseudoranges. Especially for multi-constellation receivers that can process BDS signals, the BDS GEO full measurements are acquired earlier due to the fast data rate from such satellites which will result in a faster first position fix. Hence, the proposed method is suitable for receivers processing signals from BDS and other system such as GPS.

In future, verification of cases with other constellations such as GLONASS and GALILEO should be done. Besides, how to incorporate fractional measurements from two or more constellations into this algorithm can be investigated.